# Functional surface formation by efficient laser ablation using single-pulse and burst-modes

Andrius Žemaitis, Mantas Gaidys, Justinas Mikšys, Paulius Gečys, Mindaugas Gedvilas[*]
Department of Laser Technologies (LTS), Center for Physical Sciences and Technology (FTMC), Savanorių Ave. 231, LT-02300 Vilnius, Lithuania

**ABSTRACT**

Surfaces inspired by nature and their replication find great interest in science, technology, and medicine due to their unique functional properties. This research aimed to develop an efficient laser milling technology using single-pulse- and burst-modes of irradiation to replicate bio-inspired structures over large areas at high speed. The ability to form the trapezoidal-riblets inspired by shark skin at high production speeds while maintaining the lowest possible surface roughness was demonstrated.

## 1. INTRODUCTION

Ultrafast lasers have been wildly applied in high-throughput microfabrication and patterning of metals[1–45], semiconductors[6,7], and insulators[8–11] in science[12–18], technology[19–27], and medicine[28,29]. However, their industrial applications are restricted by the low laser machining rates and the high price of ultrashort laser irradiation sources. The theoretical investigation confirmed by experimental data which clarified the maximizing the ablated volume per laser pulse by the controlling beam radius on the sample and its related laser fluence was conducted more than a decade ago[30]. The state-of-the-art scientific investigation of laser ablation efficiency suggested the ratio of laser peak fluence to the ablation threshold close to $e^2 \approx 7.39$ for the maximum material ejection rate and the most efficient ablation point[30,31]. Since then, laser ablation efficiency has been investigated in numerous papers in independent groups[32–41]. The model takes into account the transverse laser intensity distribution has a Gaussian profile, ablation occurs after fluence exceeds the ablation threshold, and ablated depth per pulse is growing logarithmically with the laser fluence on the sample. The ablated dimple has the shape of a paraboloid with the maximum possible ablated volume obtained for the applied peak laser energy density 7.39 times larger than the threshold fluence. This theoretical model has a perfect agreement with the majority of experiments[42–47]. However, the presented model failed when the laser beam is scanned along the sample surface and the trench is ablated instead of the dimple achieved by the stationary beam. Model equations predict an optimal speed of beam scanning on the target material equal to zero for the maximal ablated volume per pulse. This result is in big contradiction with the experimental data, because dimple depth saturates when a large number of pulses are applied on a single spot, and zero ablation efficiency is achieved for stationary beam ablation[48]. The maximum ablated volume per pulse for scanning beam processing and trench ablation is achieved for a non-zero scanning speed[48]. Thus, selecting beam scanning speed for the most efficient laser ablation point is still a tricky question experimentally and theoretically. Moreover, the scientific papers dedicated to the ablation efficiency were conducted only to a dimple ablation or hole drilling with stationary[49–51] and trench ablation[30,52] with scanned beams, respectively. In the multi-pulse ablation, however, the threshold decreases with an increasing number of pulses per spot which is called the incubation phenomenon and it was experimentally proven by numerous experimental investigations[32,41,53–56]. On the other hand, the depth of the ablated hole (dimple) saturates and ablated depth per pulse reaches zero values after a certain number of pulses are applied[57–59]. However, there are a limited amount of papers exploring the experimental and especially theoretical insights of laser milling. We have gained important results, finding the numerical approach to predict an optimal laser processing parameter set for the most efficient ablation point, by creating a new and competitive theoretical model for rectangular cavity milling by ultrashort laser[48]. The model considers the ablation depth saturation as well as the decrease in the ablation threshold with an increasing number of laser pulses. The modeling results coincided well with the experiment data. Also, we have demonstrated that the laser milled surface has minimal roughness with mirror-like polishing of and the best processing quality was performed using similar laser processing parameters that were used to get the ablation with the highest throughput[60]. The ability to form functional surfaces of the shark-skin on two-and-a-half-dimensional (2.5D) cylindrical surfaces[61]. It was also demonstrated that the ablation of pre-heated polytetrafluoroethylene (PTFE) was more efficient than the conventional laser ablation in normal conditions[62]. The possibility to replicate the great variety of bio-inspired surfaces as shark skin-like structures, drag-reducing trapezoidal-riblet structures, and fish-scale like structures or other surfaces on complex 2.5D surfaces at high-throughput keeping the processing quality to perfection has been shown[61]. Moreover, to the best of our knowledge, we have reached the highest laser milling ablation efficiency for copper reported in the literature using a state-of-the-art bi-burst (GHz burst in the MHz burst) femtosecond laser[63].

[*] mgedvilas@ftmc.lt; lts-ftmc.lt

Bio-inspired functional surfaces and their mimicking are getting a large amount of interest in technology, science, and medicine due to exceptional functional properties[64–67]. The most wildly investigated are the surfaces that mimic the shark-skin like structures which decrease friction with gases and liquids and also has antibacterial functionality[68–73]. The efficient laser milling is the perfect technique that enables replicating bio-inspired with unique functions at high fabrication rates[48].

This investigation aimed to develop high-throughput laser-ablation technology with maximum possible material removal rate at the same time achieving the processing quality with mirror polishing finish to mimic shark-skin-like surfaces at high fabrication rate.

## 2. EXPERIMENT

The experimental setup equipped with laser and galvanometer scanner (Figure 1. (a)) was used laser milling experiments with bidirectional beam patterning (Figure 1. (b)).

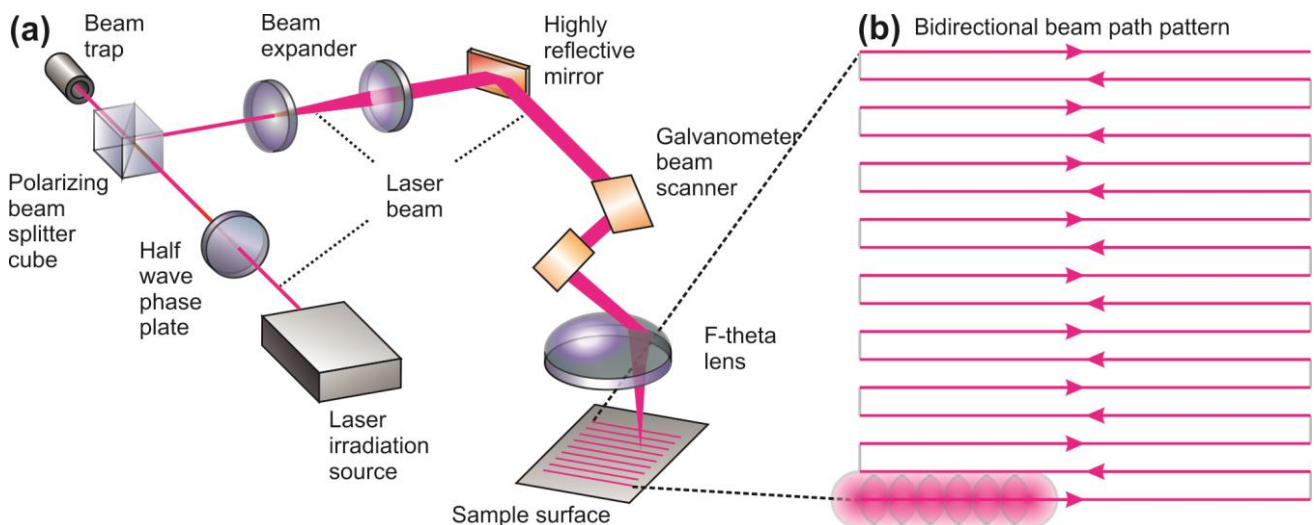

Figure 1. Principal setup of the laser ablation experiment and Gaussian beam scanning path. (a) Schematics of the laser milling stage equipped with an ultrafast laser, half-wavelength phase plate, polarizing beam splitter cube for power variation, laser beam trap, beam expander made of convex and concave lenses, high reflection mirror, galvanometer scanner with controllable galvanometric motors, and the f-theta lens for beam focusing on the surface of the sample. (b) Scheme of bidirectional scanned beam path with shown scanning direction and overlapped beam spots.

The scheme of the laser milling experiment is given in Figure 1. (a). The Gaussian beam scanning on the surface of the target material was implemented by utilizing a scanner (Scangine 14, Scanlab). The positioning of the beam on the surface of the sample at speeds up to 0.3 m/s with the bidirectional snake-like beam scanning path. The f-theta lens with a focusing distance of 80 mm was used for the Gaussian beam guiding on the copper surface. The beam scanning velocity and Gaussian beam spot radius were controlled during the tests. Beam size was controlled by changing the sample $z$ distance in respect to focus. The controlled focused beam radiuses resulted in controlled laser fluences. The beam scanning path was a bidirectional snake-like trajectory made of parallel linear scanning of overlapping pulses Figure 1. (b). The overlapping distance between lines was the same as of overlapping pulses for simplicity. The rectangular-shaped cavities were milled by using laser irradiation with depth depending on the fabrication procedure and laser settings, the laser milling efficiencies were characterized from profiles by a stylus profilometer (Dektak 150, Veeco). The laser milling rate evaluation was performed using maximal available laser power to achieve the highest possible ablation rate. The three different lasers with the parameters provided in Table 1 were used in our current and previous laser milling experiments[47,48,63].

Table 1. Lasers and main processing parameters of laser milling efficiencies of copper.

| Laser model, manufacturer | Baltic HP, Ekspla Ltd | Atlantic HE, Ekspla Ltd | Carbide, Light Conversion Ltd | Pharos, Light Conversion Ltd |
|---|---|---|---|---|
| Purse duration | 10 ns | 10 ps | 10 ps | 210 fs |
| Wavelength | 1064 nm | 1064 nm | 1030 nm | 1030 nm |
| Laser power | 12 W | 13 W | 36 W | 7.3 W |
| Repetition rate | 100 kHz | 100 kHz | 300 kHz | 100 kHz |
| Irradiation mode | Pulsed | Pulsed | Burst, 64.5 MHz | Burst, 64.7 MHz |
| Milling rate | 1.4 mm$^3$/min | 2.0 mm$^3$/min | 10.4 mm$^3$/min | 2.48 mm$^3$/min |
| Milling efficiency | 1.95 μm$^3$/μJ | 2.56 μm$^3$/μJ | 4.84 μm$^3$/μJ | 5.66 μm$^3$/μJ |
| Reference | Current work | 48 | 47 | 63 |

The highest copper milling efficiencies were achieved by using lasers with built-in picosecond- and femtosecond-burst irradiation modes (Table 1). However, it is not necessary to have a burst laser. The pulse-pairs[9,74] and burst modes[75,76] can be generated by using external setups depicted in Figure 2.

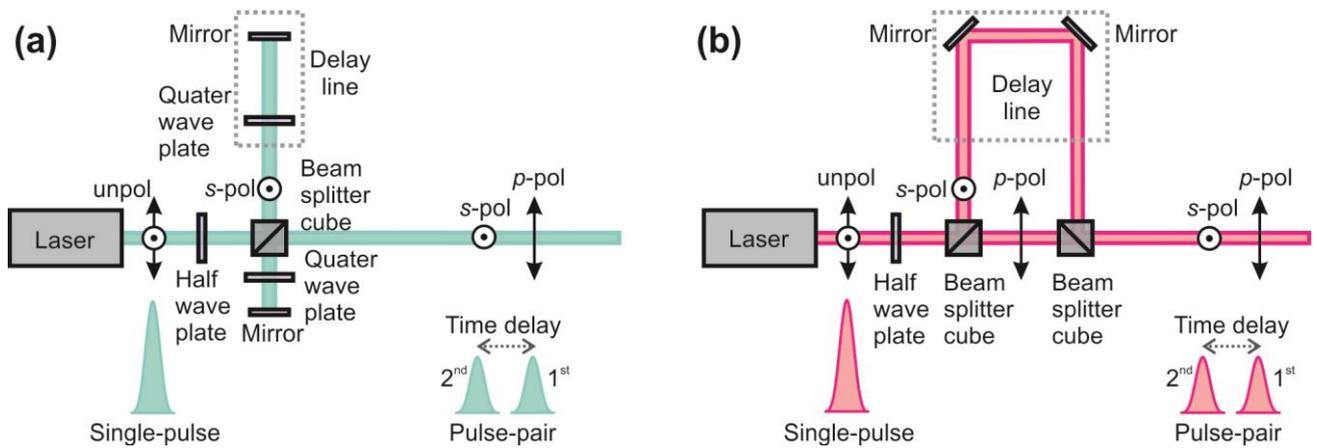

Figure 2. External setups for the pulse pair and burst generation. (a) The pulse-pair generation setup consisting of a half-wave phase plate, polarizing beam splitter cube, two quarter-wave phase plates, delay line, and two mirrors. (b) The pulse-pair generation setup consisting of one half-wave phase plate, two polarizing beam splitter cubes, a delay line, and two mirrors.

# 3. RESULTS

The copper metal samples (CW004A, Ekstremalė) with the width, length, and depth of $50 \times 50 \times 5$ mm$^3$, purity of 99.9%, and the mirror-like finish of $R_a < 0.1$ μm was used in the laser milling experiments. The laser milled surface was characterized by using a scanning electron microscope (SEM) (JSM-6490LV, JEOL). The SEM micrographs of an array of laser-ablated rectangular shaped cavities are shown in Figure 3.

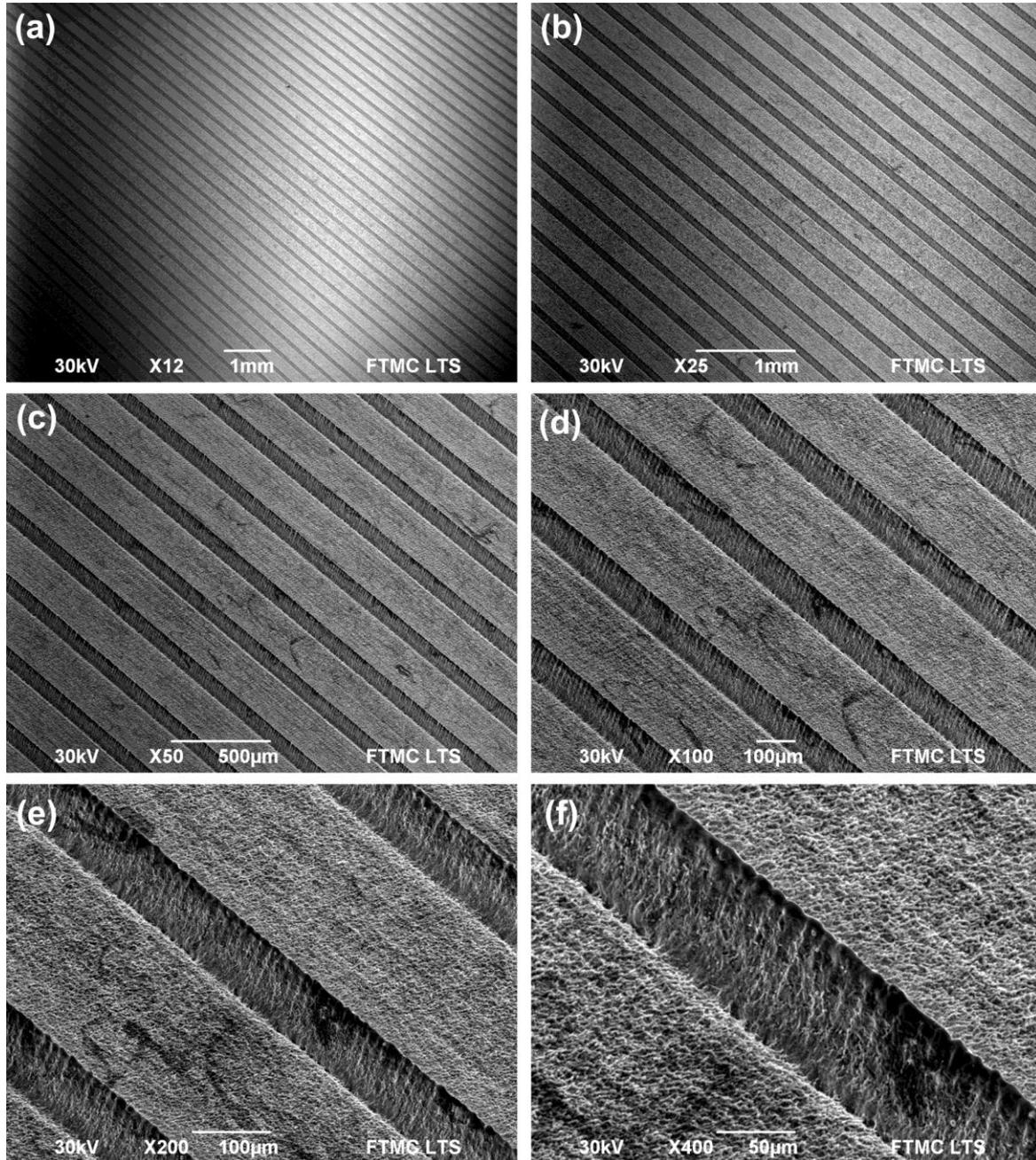

Figure 3. Trapezoidal-riblets laser-milled in the copper surface using nanosecond laser (Baltic, Ekspla). SEM images were taken at a lilt angle of 70° and magnification factors of (a) X12, (b) X25, (c) X50, (d) X100, (e) X200, and (f) X400. Laser pulse duration 10 ns, irradiation wavelength 1064 nm, pulse repetition rate 100 kHz, average laser power 12 W, copper removal rate of 1.4 mm$^3$/min, fabrication speed ~7 mm$^2$/min, milling efficiency 1.95 μm$^3$/μJ, blade period 0.4 mm, blade height 0.2 mm.

The laser milled surface was characterized by an optical profilometer (S neox, Sensofar) depicted in Figure 4.

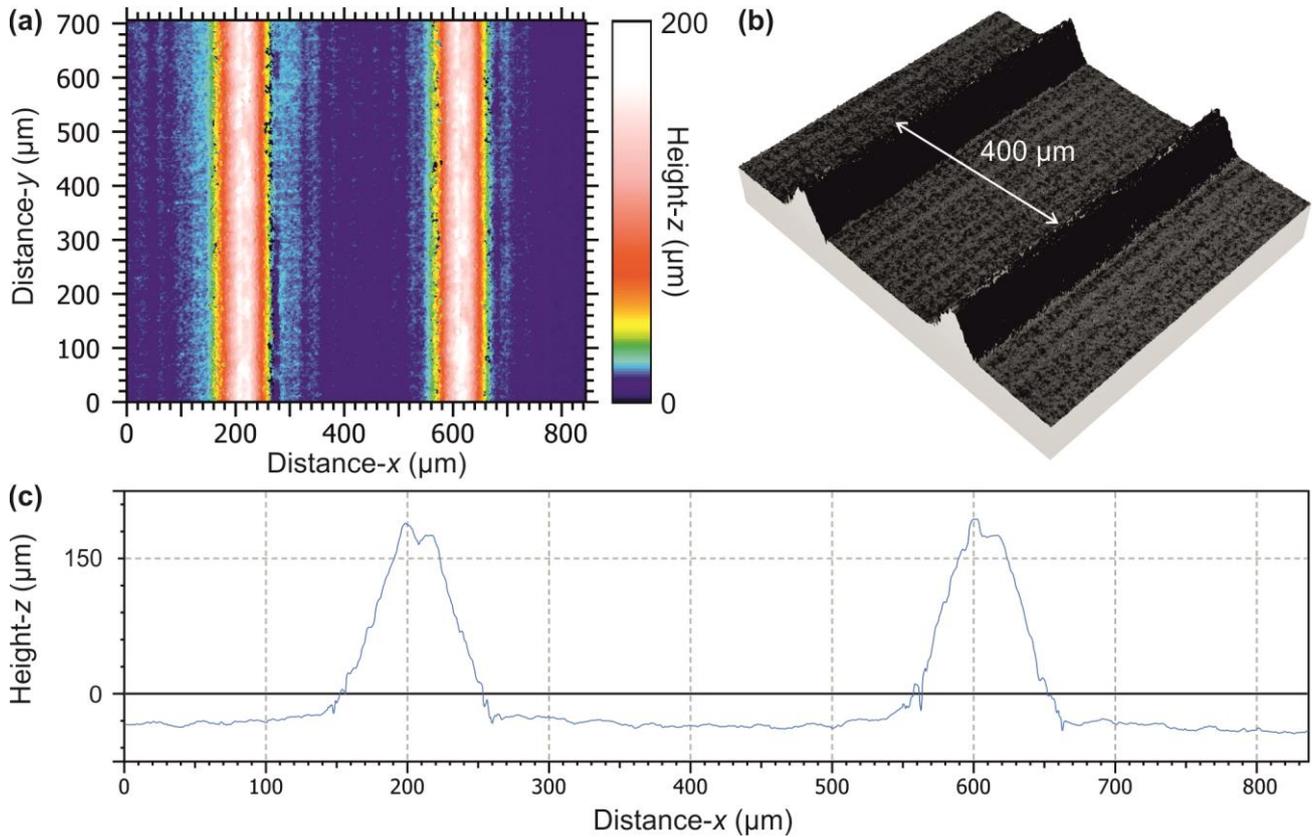

Figure 4. Profiles of trapezoidal-riblets laser-milled in the copper surface using nanosecond laser (Baltic, Ekspla). Laser pulse duration 10 ns, irradiation wavelength 1064 nm, pulse repetition rate 100 kHz, average laser power 12 W, copper removal rate of 1.4 mm$^3$/min, fabrication speed ~7 mm$^2$/min, milling efficiency 1.95 µm$^3$/µJ, blade period 0.4 mm, blade height 0.2 mm.

## 4. DISCUSSION

The scientific motivation of this investigation was previous papers that published more than a decade before theoretically predicting laser-ablated dimple shape of paraboloid and theoretically predicting the laser spot size dependent fluence for the laser ablation with the highest possible efficiency. However, the reduction of the ablation threshold for multi-pulse and ablation depth saturation with the increasing number of laser pulses per irradiated spot was not incorporated in that model[31]. In our recent work, the analytical model and its related numerical simulation for laser milling taking into account the reduction of the ablation threshold ablation depth saturation for multi-pulse ablation are declared[48]. Laser milling with nanosecond/picosecond pulses and femtosecond burst mode has been used for copper samples. The laser fabrication process was controlled by controlling beam scanning speed, line hatching distance, and fluence change by variation beam radius on the sample. The maximal possible copper removal rate was achieved experimentally. The maximum ablation efficiency 5.66 µm$^3$/µJ has been achieved for the femtosecond burst laser (Table 1). For the picosecond burst the milling efficiency of 4.84 µm$^3$/µJ has been achieved (Table 1). The picosecond single pulses provided smaller milling efficiency of 2.56 µm$^3$/µJ (Table 1). The smallest milling efficiency of 1.95 µm$^3$/µJ was achieved using a nanosecond laser (Table 1). The efficient laser milling procedure was used to replicate bio-inspired shark-skin-like trapezoidal-riblets with unique functions at high fabrication rates (Figure 3 and Figure 4).

## 5. CONCLUSIONS

To summarize, the optimized and precisely controlled laser milling technology is a versatile tool that enables replicating the variety of bio-inspired functional surfaces that has been evolving in nature for millions of years. The constant growth of affordability and reliability of ultra-short lasers, together with the increase of output optical power per price ratio in connection with the research and development of laser milling and polishing technologies opens up new opportunities for the real manufacturing of nature-inspired functional surfaces starting from small-batch laboratory production to the large-scale industrial applications. The presented theoretical and experimental data of laser milling and its direct application lead to the technological conclusion, that laser irradiation with beam scanning control is a flexible technique for replication of bio-inspired structures naturally evolved over millions of years.

## FUNDING

Agency for Science, Innovation and Technology, 01.2.2-MITA-K-702-06-0012.